\documentclass[twocolumn,showpacs,preprintnumbers,amsmath,amssymb]{revtex4}
\usepackage{graphicx}
\usepackage{dcolumn}
\usepackage{bm}
\usepackage{comment}
\usepackage{rotating}
\usepackage{longtable}
\usepackage{float}
\usepackage{eucal}
\usepackage{csquotes} 

\begin{document}

\title{\centering {Abrupt phase change of the core rotation in the $^{143}$Sm nucleus}}

\author{S. Rajbanshi$^{1}$}
\email{subhphy@gmail.com}
\author{R. Raut$^{2}$}
\author{H. Pai$^{3}$}
\author{Sajad Ali$^{3}$}
\author{A. Goswami$^{3}$}
\author{S. Bhattacharyya$^{4}$}
\author{G. Gangopadhyay$^{6}$}
\author{G. Mukherjee$^{4}$}
\author{S. Muralithar$^5$}
\author{M. Kumar Raju$^{7}$}
\author{P. Singh$^{8}$}
\author{R. P. Singh$^5$}
\author{R. K. Bhowmik$^{5}$}

\affiliation{$^1$Department of Physics, Dum Dum Motijheel College, Kolkata 700074, India}
\affiliation{$^2$UGC-DAE-Consortium for Scientific Research, Kolkata 700098, India} 
\affiliation{$^3$Saha Institute of Nuclear Physics, 1/AF, Bidhannagar, Kolkata 700064, India}
\affiliation{$^4$Variable Energy Cyclotron Center, Kolkata 700064, India}
\affiliation{$^5$Inter University Accelerator Centre, Aruna Asaf Ali Marg, New Delhi 110067, India}
\affiliation{$^6$Department of Physics, University of Calcutta, Kolkata 700009, India}
\affiliation{$^7$Nuclear Physics Department, Andhra University, Visakhapatnam 530003, India}
\affiliation{$^8$Indian Institute of Technology, Kharagpur 721302, India}

\date{\today}

\begin{abstract}

Dipole sequences in the $^{143}$Sm nucleus have been investigated via the $^{124}$Sn ($^{24}$Mg, 5n) reaction at $E_{lab}$ = 107 MeV using the Indian National Gamma Array (INGA). The spin-parity of the associated levels have been firmly established from the spectroscopic measurement. Level lifetimes of several levels in the dipole bands have been measured using the Doppler Shift Attenuation Method. The decreasing trend of the measured $B(M1)$ and $B(E2)$ transition strengths in one of the sequence (DB I) spells out its origin as Magnetic Rotation (MR). The trends of B(M1) and B(E2) in DB I are reproduced well in the theoretical calculations using the Shears mechanism with the Principal Axis Cranking (SPAC) model. However, the calculations fail to reproduce the sharp rise in the $B(M1)$/$B(E2)$ ratio 
at the highest spins in DB I and the same has been interpreted from the decreasing of the core rotation along the sequence. The experimental observations along with the the theoretical calculations for the second dipole band (DB II), indicate that the core rotation, rather than the shears mechanism, is being favored for angular momentum generation. This represents a unique observation of forking of the shears band DB I from an abrupt phase change of the core from spherical into the deformed one. \\  
 
\end{abstract}


\maketitle

The observation of band-like structure in weakly deformed nuclei near shell closures has long been of interest in nuclear structure studies. These nuclei, generally, do not have an asymmetry in mass distribution, like the mid-shell ones which break the rotational symmetry and consequently exhibit regular sequences characterized by E2 intraband transitions in their level structure \cite{abohr, defor1, defor2}. In the spherical or weakly deformed nuclei in the vicinity of the shell closures, the current distribution of few nucleons (particles and/or holes) outside the core crucially contribute in the generation of angular momentum and several exotic phenomena have been observed in such systems \cite{frauen1}. \\

The \enquote{shears mechanism}, following a detailed experimental and theoretical investigations \cite{ajsim, hubel,amita}, has been identified as one of such phenomena associated with the generation of angular momentum in nuclei with few particles and holes outside a weakly deformed core. The manifestation of the mechanism has been established as a regular sequence of levels connected by strong M1 intraband transitions while the cross-over E2 transitions are either weak or absent in conformity with the weak deformation characteristics in these nuclei. The perpendicular component of magnetic dipole moment of the valence nucleons rotates about the total angular momentum axis and hence identified as Magnetic Rotational (MR) band. The experimental signature of the MR band structure is a decreasing trend of the 
B(M1) and B(E2) values with increasing spin and the same has been well interpreted within the theoretical framework invoked for the purpose \cite{frauen1, rmcla1}. As is now understood, particles and holes outside the weakly deformed core can form an arrangement of two perpendicularly aligned angular momentum vectors (blades) that represent the band head configuration of the MR band. Angular momentum along the band is generated by closing-in of the two blades, much like that of a pair of shears that renders the name shears band to the resulting sequence. Such an arrangement of particle and hole angular momentum leads to a large transverse component of the magnetic dipole moment (see, for example, Fig. 4 of Ref. \cite{frauen1}) that explains the strong intraband M1 transitions observed in MR bands. This transverse component of the magnetic dipole moment decreases in magnitude with the closing-in of the angular momentum vectors thus explaining the decreasing trend of the B(M1) values along the band. Eventually the two blades are completely aligned to the total angular momentum vector while generating the maximum possible spin corresponding to the particle-hole configuration assigned to the 
respective MR band \cite{frauen1}.  \\

Though the nuclei in the vicinity of the shell closures do not have a stable deformation in the ground state, the excited multi-quasiparticle configurations involving high-$j$ deformation driving orbitals, associated with MR bands, can extend a polarizing effect on the core leading to a weak deformation of the latter, manifested in the weak cross-over E2 transitions in these bands. In fact the total angular momentum of the constituent states in MR bands has contributions not only from the aligning angular momentum vectors generated by particles and holes but also from the rotation of the weakly deformed core. As the current distributions of the particles and the holes become more symmetric along the MR band, their polarizing effect correspondingly decreases and the same is indicated by the diminishing B(E2) strengths with increasing spin in the MR bands. Interestingly, however, the B(M1)/B(E2) ratio also exhibits a decreasing trend along the band thus implying an increasing core contribution along the MR sequence. This is appropriately interpreted by the Shears mechanism with Principal Axis Craking (SPAC) model that has been successfully applied to the MR bands observed in $^{139}$Sm, $^{141}$Eu and $^{142}$Gd \cite{pasern, podsvi, pasern1}. Quite uniquely, contrary to the aforementioned generality in the trend of the B(M1)/B(E2) ratio along the MR bands, a smooth increase in this ratio has been observed in one of the dipole sequences in the $^{144}$Dy nucleus \cite{mgproc}. This band was ascribed to the MR phenomenon but no conclusive explanation could be presented for the idiosyncracy.  In the wake of such observations, a probe into the role of core rotation in the MR phenomenon, its contribution in the generation of angular momentum therein and its evolution along the band is warranted. \\  

The present paper reports such a study in the $^{143}$Sm ($Z = 62, N = 81$) nucleus. The nucleus was previously studied by Raut {\it{et al.}} \cite{rraut} using heavy-ion induced fusion-evaporation reaction and an array of six Compton suppressed Clover detectors as the detection system. A dipole band, with a band-head energy of $\sim$ 8.6 MeV and tentative spin-parity 43/2$^-$ was reported therein. The band was tentatively interpreted in the light of the MR phenomenon but, in the absence of any lifetime measurements, no conclusive evidence for the proposition could be obtained. The current work re-investigates the dipole structure (DB I) in $^{143}$Sm using an array of higher efficiency provides confirmation to the MR mechanism associated with the sequence and reports a new dipole band (DB II) feeding DB I, at one of its intermediate states. Such bifurcation in the level structure is understood to embody a novel behaviour of this nucleus in the generation of angular momentum at the highest excitations. \\

The dipole structures in $^{143}$Sm were populated in the present study using $^{124}$Sn($^{24}Mg,5n$) reaction at $E_{lab}$ = 107 MeV. The beam was obtained from the 15 UD pelletron at the Inter University Accelerator Center (IUAC), New Delhi. The target was 0.8 mg/cm$^2$ enriched (99.9\%) $^{124}$Sn evaporated on a 13 mg/cm$^2$ gold backing. The residues were produced with velocity $\beta \sim 1.6$\%. The de-exciting $\gamma$ transitions were detected using the Indian National Gamma Array (INGA) which consisted of eighteen Compton suppressed clover detectors arranged in four different angles 90$^{\circ}$(6), 123$^{\circ}$(4), 148$^{\circ}$(4) and 57$^{\circ}$(4) with respect to the beam axis (the number in the parenthesis is the detector numbers at the respective angles) \cite{smurali}. About 8$\times$10$^{8}$ two and higher fold $\gamma$-$\gamma$ coincidence events were collected in list mode format. The acquired data was sorted into different symmetric and angle dependent $E_{\gamma}$ - $E_{\gamma}$ matrices with INGASORT \cite{ingasort} and RADWARE \cite{radford1, radford2} codes which were also used for the subsequent analysis. The multipolarities and the electromagnetic characters of the observed $\gamma$-ray transitions, for assigning the spin-parity of the levels, were determined from the measurements of the ratio for Directional Correlation from Oriented ($R_{DCO}$) state \cite{kramer, kabadi}, the asymmetry in linear polarization ($\Delta$) \cite{staro2, droste2, deng2, jones2} and the mixing ratio ($\delta$) \cite{tyamaza, edmawsu, macias}. The experimental details and data analysis procedures have been described in details in Ref. \cite{rajban, rajban1}. \\

\begin{figure}[t]
\centering
\setlength{\unitlength}{0.05\textwidth}
\begin{picture}(10,9.5)
\put(-3.40,-1.8){\includegraphics[width=0.80\textwidth, angle = 0]{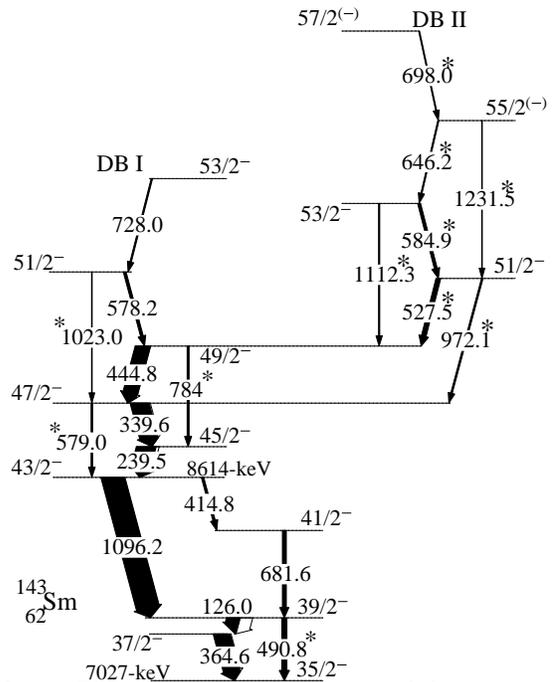}}
\end{picture}
\caption{\label{levsch} The proposed level structure of the negative parity states above the 7027-keV 35/2$^{-}$ state in $^{143}$Sm. The newly observed transitions are marked by an asterisk.}
\end{figure}

The proposed dipole bands in $^{143}$Sm, as observed in the present work, is illustrated in Fig. \ref{levsch}. In the previous study \cite{rraut} the spin-parity assignments of the associated levels were tentative owing to the large uncertainities or impossibilities in the $R_{DCO}$ and polarization asymmetry measurements of the transitions, such as 126 keV, leading to the dipole band. In the current investigation, a 490.8 keV transition has been observed de-exciting the same 7517 keV level as the 126 keV transition. The $R_{DCO}$ and polarization asymmetry values of this 490.8 keV $\gamma$-ray establish its E2 character, following which the spin-parity of the 7517 keV level could be confirmed to be 39/2$^-$, upholding the previous (tentative) assignment by Raut {\it{et al.}}. Further, the $R_{DCO}$ and polarization asymmetry values for the 1096.2, 414.8 and 681.6 keV transitions are consistent with the previous assignments and consequently the spin-parity of the 8614 keV state has been conclusively assigned as 43/2$^-$. \\

Raut {\it{et al.}} had reported a dipole sequence in the $^{143}$Sm nucleus starting at 8614 keV level and extending upto a spin of 57/2$^-$. The $R_{DCO}$, polarization asymmetry and multipole mixing ratio of the first four member transitions of this band, 239.5, 339.6, 444.8 and 578.2 keV, determined in the present study establish their predominantly M1 nature in compliance with the earlier assignments and confirming the spin-parity of the sequence upto 51/2$^-$. Above this state, Raut {\it{et al.}} had assigned three $M1$ transitions, 602, 728 and 706 keV, to the dipole band. The present work, carried out using a substantially more sensitive detection setup, has confirmed the 728.0 keV transition as a member of the dipole band above 51/2$^-$ state while the 602.0 and 706.0 keV transitions, notwithstanding their coincidence with most of the transitions in the sequence, are not part of the sequence designated as DB I in the present study. Thus the state de-excited by the 728 kev transition is the highest observed state of DB I and has been assigned a spin-parity of 53/2$^-$. We have also observed the weak cross-over transitions between 47/2$^-$$\rightarrow$43/2$^-$, 49/2$^-$$\rightarrow$45/2$^-$ and 51/2$^-$$\rightarrow$47/2$^-$ states in DB I (Fig. \ref{levsch}). The $R_{DCO}$ and polarization asymmetry measurements of these confirm their E2 character. \\   

The salient finding of the present work is the observation of a bifurcation in DB I at 49/2$^-$ leading to another sequence of levels connected by the 527.5, 584.9, 646.2 and 698.0 keV transitions. The $R_{DCO}$ and polarization asymmetry measurements of 527.5 and 584.9 keV $\gamma$-rays indicate their predominantly M1 nature. The polarization asymmetry of 646.2 and 698.0 keV transitions could not be 
determined owing to the want of statistics at these highest excitations but the $R_{DCO}$ values for these do indicate their dipole character. Following these assignments, the sequence of levels forked from DB I has been interpreted as the second dipole band, labelled as DB II in Fig. \ref{levsch}. Cross-over transitions connecting 53/2$^-$$\rightarrow$49/2$^-$ and 55/2$^-$$\rightarrow$51/2$^-$ states (Fig. 1) in DB II have also been observed with statistics that did not allow for determination of their polarization asymmetry but did qualify for the extraction of the $R_{DCO}$ values that confirm their quadrupole character. \\

\begin{table*}
\centering
\caption{\label{lifetime} The DCO ratio ($R_{DCO}$), polarization asymmetry ($\Delta$), level lifetimes ($\tau$) and side feeding lifetimes ($\tau_{sf}$) of the states and the corresponding $B(M1)$, $B(E2)$ and $B(M1)$/$B(E2)$ transitions rates for the $\gamma$ transitions of the dipole bands DB I and DB II in $^{143}$Sm.}

\begin{tabular}{ccccccccccccc}
\hline\hline
$J_{i}$      &$J_{f}$    & $E_\gamma$  & $R_{DCO}$$^{a}$ & $\Delta$ & Branching  & $\tau$    & $\tau_{sf}$ &  $B(M1)$            &     $B(E2)$    & $B(M1)/B(E2)$ \\

[ $\hbar$ ]  & [ $\hbar$ ]& [ MeV ]    &           &           &   [ \% ]   & [ ps ]    &  [ ps ]     & [ $\mu_{N}^{2}$ ]  &  [ $e^2b^2$ ]  & [ $\mu_{N}^{2}$/$e^2b^2$ ]\\

\hline

DB I \\

45/2$^{-}$   &43/2$^{-}$   & 239.5  & 0.56(3)$^{Q}$  & -0.21(16) & 100  & 1.02$^{+0.16}_{-0.12}$ & 0.20(3)$^{b}$ & 3.52$^{+0.55}_{-0.42}$ &                   &\\

47/2$^{-}$   &45/2$^{-}$   & 339.6  & 0.52(3)$^{Q}$  & -0.27(4)  & 92.3 & 0.58$^{+0.09}_{-0.08}$ & 0.21(3)$^{b}$ & 2.16$^{+0.34}_{-0.28}$ &                   & 12.70$^{+2.50}_{-2.22}$\\

             &43/2$^{-}$   & 579.0  &                &           & 7.7  &                        & &                        & 0.17$^{+0.02}_{-0.02}$ &\\

49/2$^{-}$   &47/2$^{-}$   & 444.8  & 0.48(4)$^{Q}$  & -0.12(4)  & 85.8 & 0.44$^{+0.07}_{-0.06}$ & 0.18(3)$^{b}$ & 1.20$^{+0.20}_{-0.16}$ &                   & 13.33$^{+3.70}_{-2.31}$\\

             &45/2$^{-}$   & 784.3  & 1.70(15)$^{D}$ & +0.32(11) & 14.2 &                        & &                        & 0.09$^{+0.02}_{-0.01}$ &\\

51/2$^{-}$   &49/2$^{-}$   & 578.2  & 1.27(10)$^{D}$ & -0.08(3)  & 85.4 & 0.41$^{+0.07}_{-0.06}$ & 0.16(4)$^{c}$ & 0.61$^{+0.10}_{-0.09}$ &                   & 20.33$^{+4.91}_{-4.20}$\\

             &47/2$^{-}$   & 1023.0 &                &  +0.11(3) & 14.6 &                        & &                        & 0.03(1) &\\

53/2$^{-}$   &51/2$^{-}$   & 728.0  & 1.09(9)$^{D}$  &  -0.04(2) &  100  & 0.78$\downarrow$      & 0.13(3)$^{c}$ & 0.18$\uparrow$         &                   & 46.78$\uparrow$$^{d}$\\
\\
DB II \\

51/2$^{-}$  &49/2$^{-}$   & 527.5  & 0.45(4)$^{Q}$  & -0.26(8) & 74.8 & 0.78$^{+0.12}_{-0.09}$ & 0.15(3)$^{b}$ & 0.36$^{+0.06}_{-0.04}$ &             & 12.00$^{+2.61}_{-1.96}$\\

            &47/2$^{-}$   & 972.1  & 1.81(23)$^{D}$  &         & 25.2 &                        &   &                        & 0.03(1)     &\\

53/2$^{-}$  &51/2$^{-}$  & 584.9  & 0.42(4)$^{Q}$  & -0.12(5) & 73.6 & 0.57$^{+0.09}_{-0.07}$ & 0.16(4)$^{b}$ & 0.35$^{+0.06}_{-0.04}$ &             & 17.50$^{+3.91}_{-3.04}$\\

            &49/2$^{-}$   & 1112.3 & 2.09(21)$^{D}$ &         &  26.4 &                        & &                        & 0.02(1)     &\\

55/2$^{(-)}$  &53/2$^{-}$  & 646.2  & 0.91(8)$^{D}$  &        &  64.3 & 0.44$^{+0.07}_{-0.06}$ & 0.11(3)$^{c}$ & 0.31$^{+0.05}_{-0.04}$ &             & 15.50$^{+3.49}_{-2.99}$\\

              &51/2$^{-}$  & 1231.5 & 1.77(20)$^{D}$ &        & 35.7 &                        & &                        & 0.02(1)     &\\

57/2$^{(-)}$  &55/2$^{(-)}$  & 698.0  & 1.28(11)$^{D}$  &     & 100  & 0.66$\downarrow$       & 0.08(3)$^{c}$  & 0.25$\uparrow$         &             & \\

\hline\hline

\multicolumn{11}{l}{The uncertainties are rounded off to the nearest value up to two decimal places.}\\

\multicolumn{11}{l}{$^{a}$The measured $R_{DCO}$ value of the $\gamma$-transitions formthe gate on quadrupole ($Q$) and dipole ($D$) transitions.}\\

\multicolumn{11}{l}{$^{b}$Determined from level lifetime analysis using both GTT and GTB spectra, as discussed in the text.}\\
\multicolumn{11}{l}{$^{c}$Determined from an extrapolation of the side feeding times measured for some of the levels indicated with b (see text).}\\

\multicolumn{11}{l}{$^{d}$The lower limit of the $B(M1)$/$B(E2)$ value has been extracted for the 53/2$^{-}$ state of DB I from the intensity measurements.}\\

\end{tabular}

\end{table*}

Observation of Doppler affected transition peaks corresponding to the $\gamma$-rays in DB I and DB II facilitated the determination of level lifetimes in these bands using the Doppler Shift Attenuation Method (DSAM). A combination of developments reported in \cite{das16} and the LINESHAPE \cite{wells, johnson} package has been used for the corresponding analysis. The stopping of the residue through the target and the backing media were simulated using Monte Carlo based codes and updated stopping powers from the SRIM database. The simulated trajectories were then used to generate velocity profiles as viewed by the detectors. These profiles were input to calculation of the expected Doppler shapes for transitions of interest. The level lifetimes were finally determined by the least square fitting of the experimental (gated) spectra with the calculated shapes. It is understood that if the spectrum is from a gate on a transition below (GTB) the transition of interest, it is necessary to incorporate the contributions from the side feeding transitions while analyzing the level lifetime. For analysis with GTB, as far as the LINESHAPE package is concerned, the side feeding is modelled with a cascade of five transitions and the moment of inertia of the sequence is typically chosen to be similar to that of the band of interest being analyzed. The same procedure has been adopted in the present work for determining the level lifetimes in DB I and DB II. The detail analysis procedure was discussed in Refs. \cite{rajban,rajban1}.  \\

The side feeding lifetime is often identified as one of the principal uncertainties in the analysis of DSAM data. However, if the spectrum is generated from a gate on a transition that is atop (GTT) the transition of interest the effect of the side feeding is eliminated. Such spectra with gates on the higher lying transitions, above the transitions of interest, are plagued with lack of statistics, and it is often impossible to use these for lifetime analysis. In the present work, level lifetimes of some of the levels, both in DB I and DB II, could actually be extracted from both GTB and GTT spectra. These were 45/2$^-$, 47/2$^-$ and 49/2$^-$ states in DB I and 51/2$^-$, 53/2$^-$ states in DB II. This analysis facilitated extraction of side feeding times required for determination of level lifetimes from GTB spectra. The lifetime of these levels were first determined using GTT spectra and then using the GTB ones, for different values of the sidefeeding times, in order to reproduce the result obtained from the GTT. In the process, the side feeding time required for the GTB based analysis, was extracted. For the remaining states, where both the GTT and the GTB techniques could not be applied and the lifetime 
had to be determined only from the GTB, the value of the side feeding time used was determined from an extrapolation of the measured values of the same from the previous analysis \cite{rschwen}. \\

\begin{figure} [t]
\centering
\setlength{\unitlength}{0.05\textwidth}
\begin{picture}(10,7.0)
\put(-0.7,-0.6){\includegraphics[width=0.57\textwidth, angle = 0]{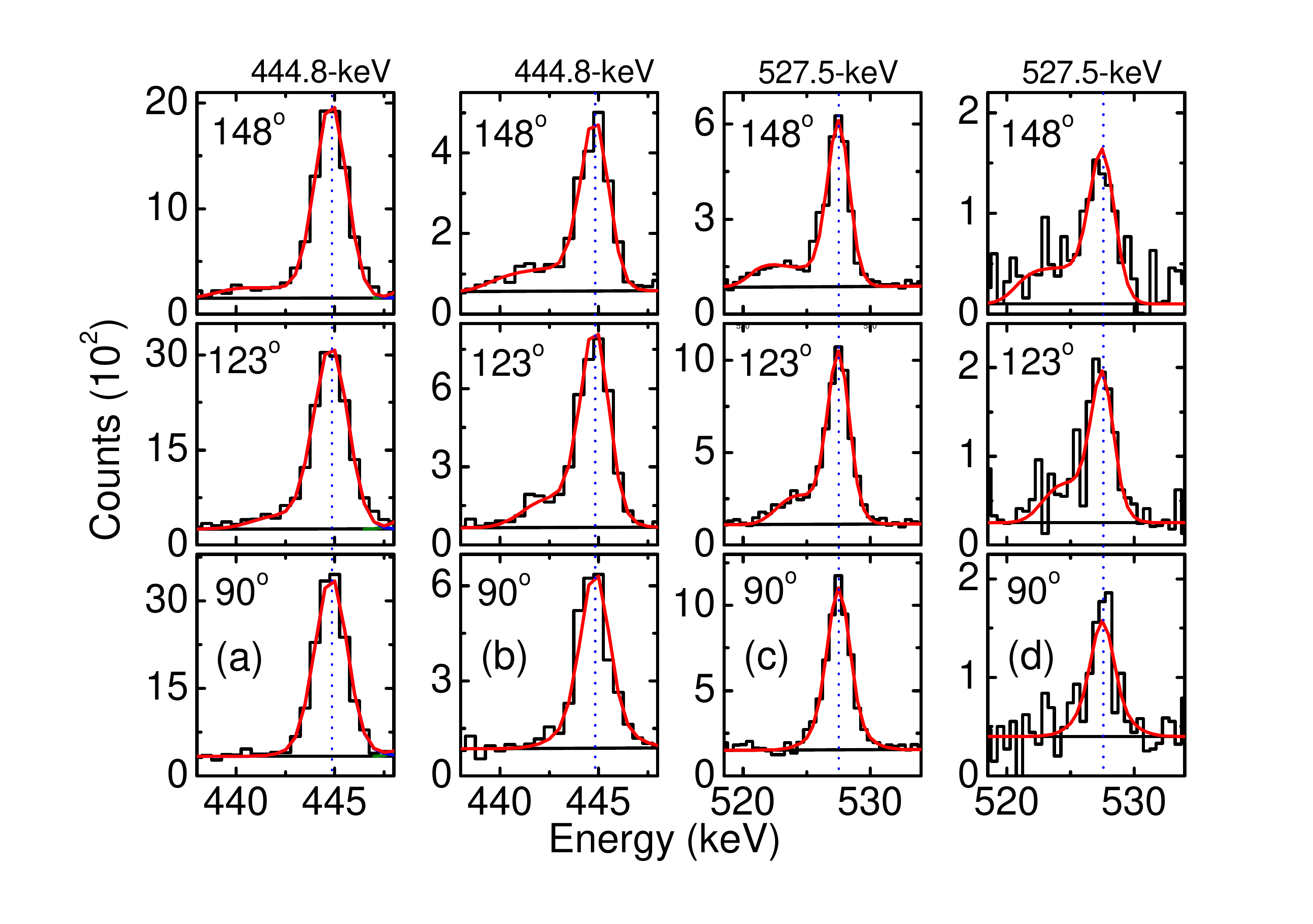}}
\end{picture}
\caption{\label{shape-bot} (Color  online) The experimental spectra along with the fitted line shapes for the $\gamma$ transitions (a) 444.8 and (c) 527.5-keV created by the \enquote{GTB} technique and, (b) 444.8 and (d) 527.5-keV created by the \enquote{GTT} technique of the bands DB I and DB II in $^{143}$Sm. The top, middle and bottom rows correspond to the shapes in the 148$^{\circ}$, 123$^{\circ}$ and 90$^{\circ}$ detectors, respectively. The obtained total line-shapes of the $\gamma$ transitions are represented by the red lines.}
\end{figure}

Table I summarizes the $R_{DCO}$, $\Delta$, and the lifetime results from the present measurements along with the B(M1) and the B(E2) values in bands DB I and DB II in $^{143}$Sm. Fig. \ref{shape-bot} illustrates typical fits to the experimental Doppler affected spectra that were simultaneously fitted at three different angles, 148$^o$, 123$^o$ and 90$^o$, to obtain lifetime results. The uncertainties on the quoted lifetimes were derived from behaviour of the $\chi^2$ value in the vicinity of the least square minimum. The uncertainties do not include the systematic contribution of the stopping powers that, owing to the use of updated and experimentally benchmarked stopping powers from the SRIM database, are expected to be $\sim$ 5\%. The absence or non-observation of the cross-over (E2) transition 53/2$^{-}$$\rightarrow$49/2$^{-}$ in DB I actually results in a large value of the $B(M1)$/$B(E2)$ ratio for the 53/2$^{-}$ state. The lower limit of this value can be estimated from the minimum intensity of the transition required for observing it in the coincidence spectrum. In the present data, we were able to observe the $\gamma$-ray transitions with minimum intensity at 1\% with respect to that of the 239.5 keV transition. Considering this to be the maximum intensity of the unobserved cross-over transition, the lower limit of the ratio of the $B(M1)$ and the $B(E2)$ values comes out to be of 46.78 $\mu_{N}^{2}$/$(eb)^{2}$, that has been recorded in Table \ref{lifetime}. \\

\begin{figure}[h]
\centering
\setlength{\unitlength}{0.05\textwidth}
\begin{picture}(10,11.0)
\put(-0.0,-0.5){\includegraphics[width=0.50\textwidth, angle = 0]{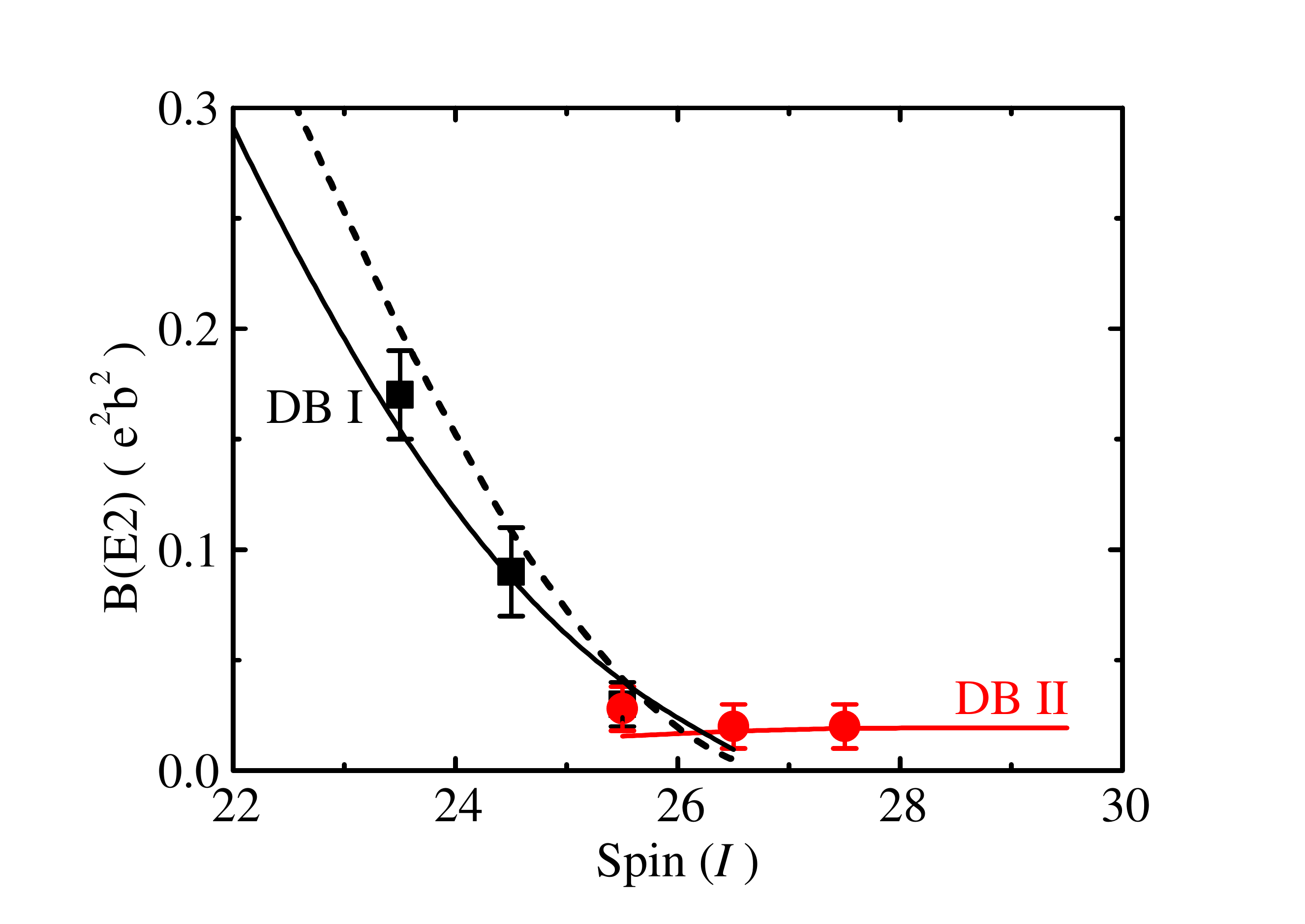}}
\put(-0.0,+4.9){\includegraphics[width=0.50\textwidth, angle = 0]{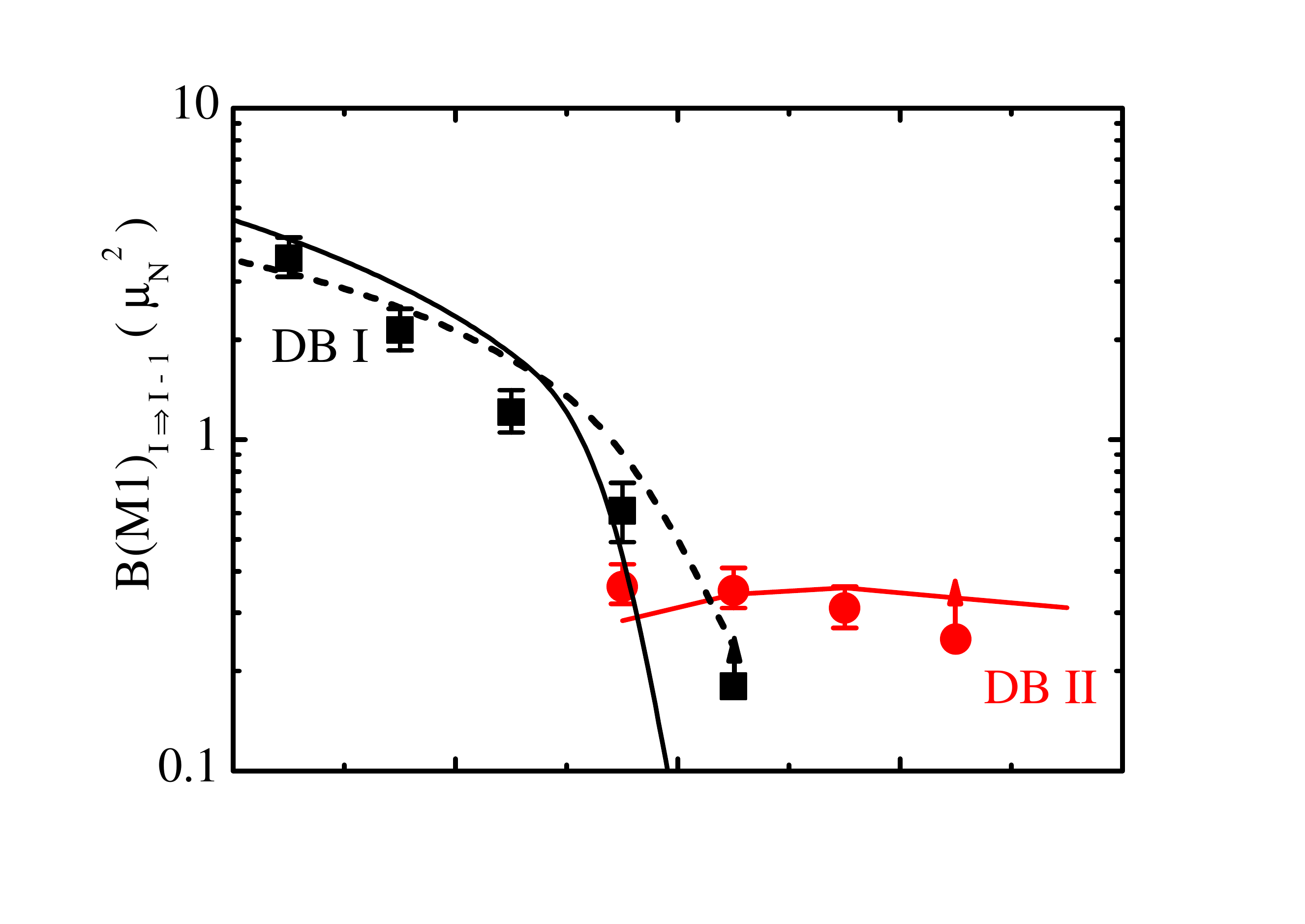}}
\put(+3.50,+1.75){\includegraphics[width=0.28\textwidth, angle = 0]{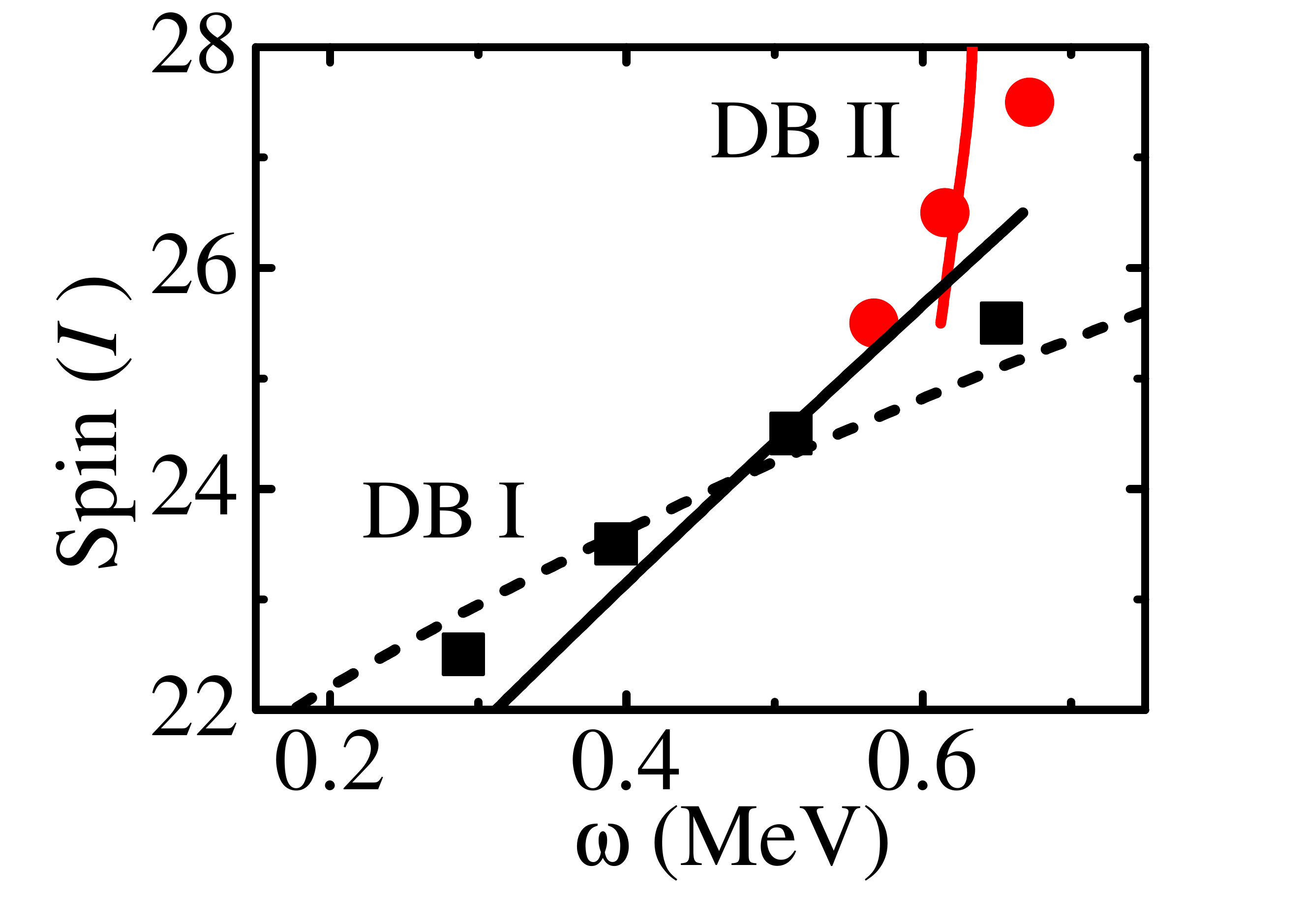}}
\put(+4.45,+7.65){\includegraphics[width=0.23\textwidth, angle = 0]{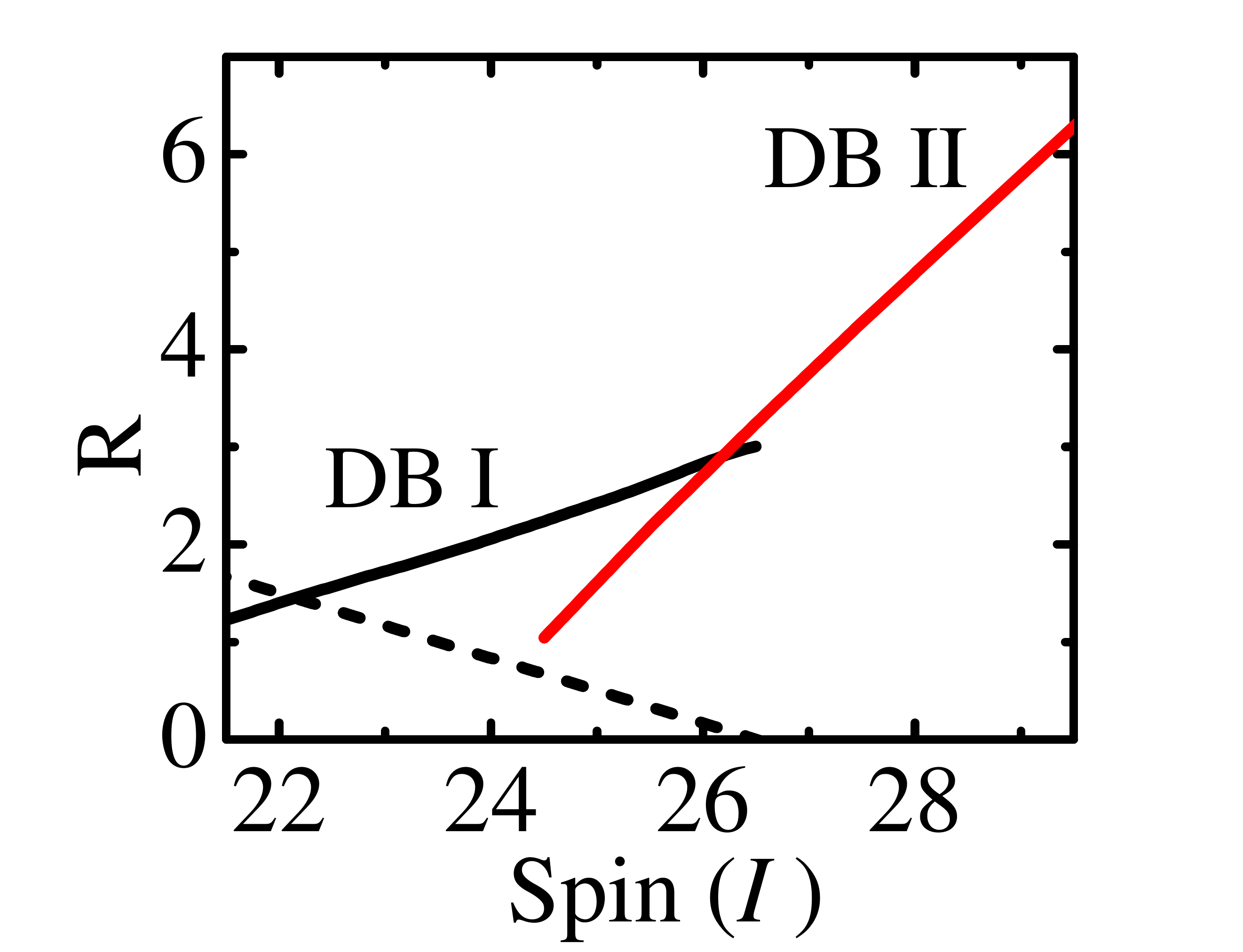}}
\put(2.5,6.7){\textbf{(a)}}
\put(2.5,1.2){\textbf{(b)}}
\end{picture}
\caption{\label{theo-cal} (Color  online) Comparison of the experimental results for the dipole bands DB I (represented by the black filled squares) and DB II (represented by the red filled solid circles) in $^{143}$Sm with the SPAC model (solid black and red lines for the DB I and DB II, respectively) and the Clark and Macchiavelli's description (black dashed lines). The $B(M1)$ and $B(E2)$ transition strengths against spin ($I$) have been depicted in (a) and (b), respectively. The variation of $R$ with spin ($I$) and the nature of spin ($I$) against the rotational frequency ($\omega$) are shown in the inset of (a) and (b), respectively.}
\end{figure}

The dipole band DB I in $^{143}$Sm is characteristically similar to the MR bands observed in the neighbouring nuclei such as $^{139, 141, 142}$Sm, $^{141,143}$Eu, and $^{142}$Gd \cite{pasern, podsvi, pasern1, rajban1, rajban, rajban3}. The transitional probabilities, B(M1) and B(E2), in DB I exhibit the characteristic decrease with increasing spin, as illustrated in Fig. \ref{theo-cal}, identical to that for the MR bands in the aforementioned nuclei in the vicinity. In fact the values of B(M1) and B(E2) in DB I are comparable to those in the other MR bands identified in this mass region. These observations uphold the proposition that the DB I sequence in the $^{143}$Sm nucleus is a MR band. \\

\begin{figure}[t]
\centering
\setlength{\unitlength}{0.05\textwidth}
\begin{picture}(10,6.3)
\put(-0.6,-1.0){\includegraphics[width=0.55\textwidth, angle = 0]{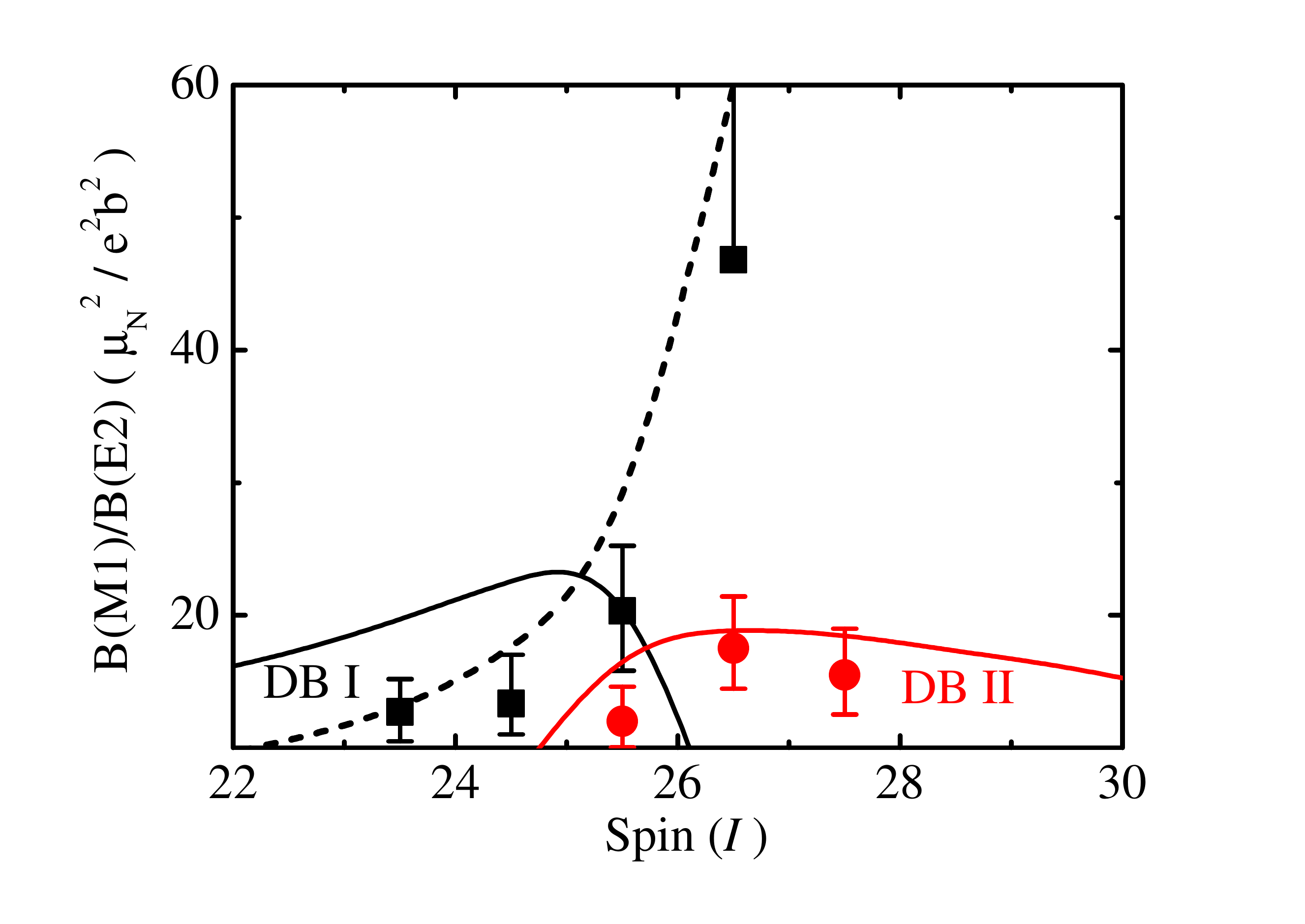}}
\put(+5.00,+1.50){\includegraphics[width=0.20\textwidth, angle = 0]{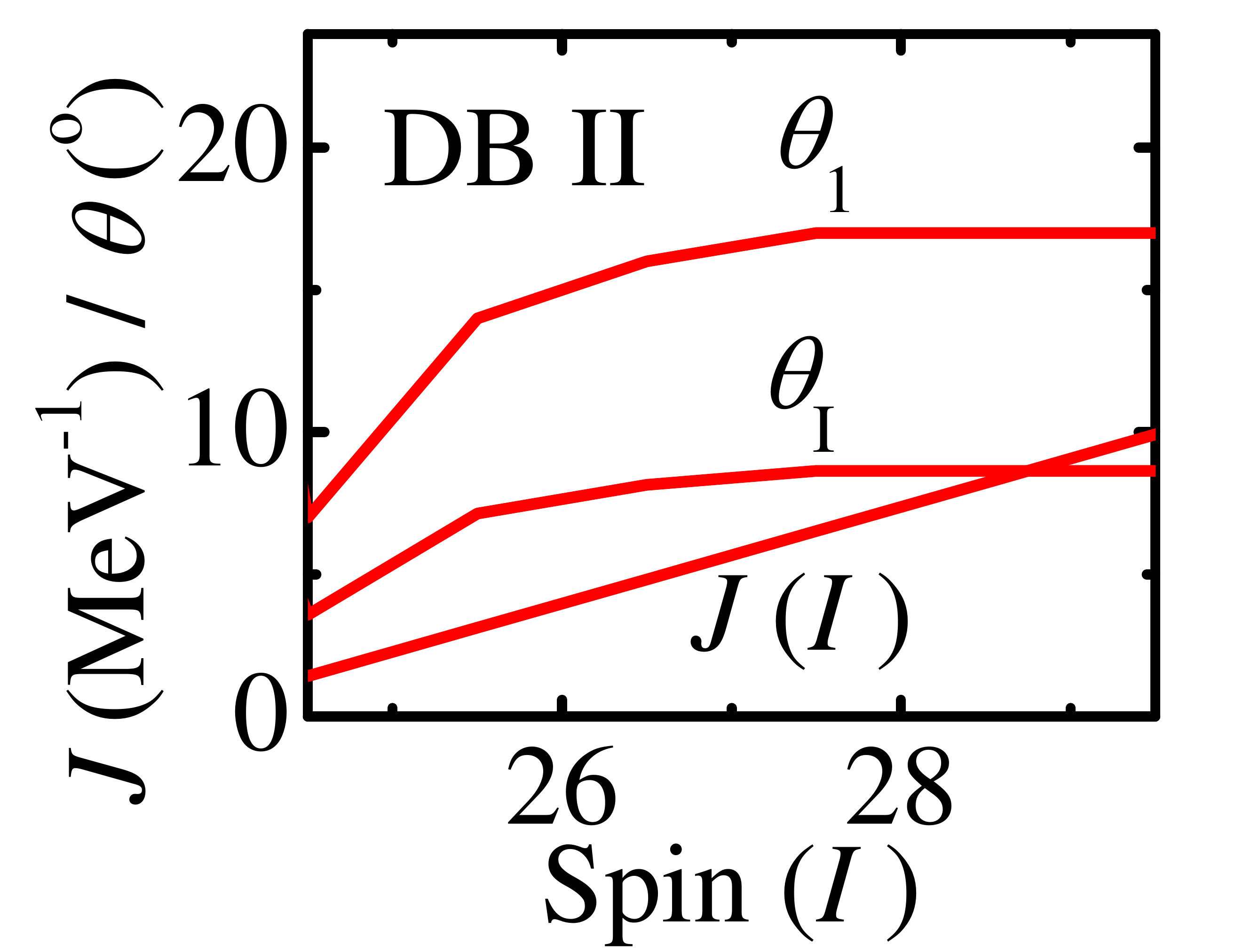}}
\put(+1.20,+3.00){\includegraphics[width=0.21\textwidth, angle = 0]{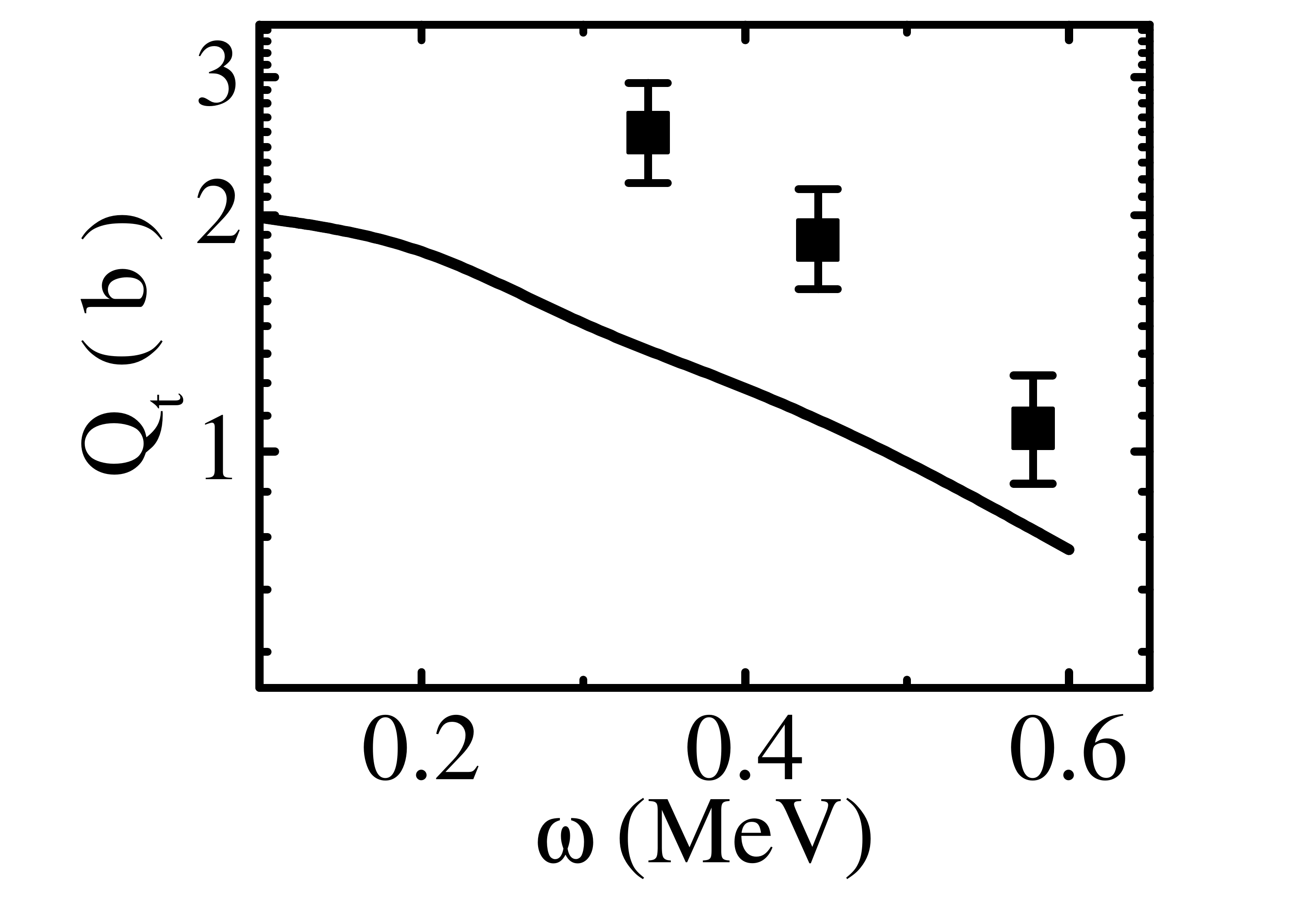}}
\put(8.1,4.1){\textbf{(b)}}
\put(2.5,4.3){\textbf{(a)}}
\end{picture}
\caption{\label{tran-bm1-be2} (Color  online) Comparison of the experimental $B(M1)$/$B(E2)$ values for the dipole bands DB I (represented by the black filled squares) and DB II (represented by the red filled solid circles) in $^{143}$Sm with the SPAC model (solid black and red lines for the DB I and DB II, respectively) and the Clark and Macchiavelli's description (black dashed lines). Inset (a) shows the expeimental transition quadrupole moment parameter $Q_{t}$ (solid black square) along with the results obtained from the TRS calculations (solid black line) for DB I. Inset (b) shows the moment of inertia of the core and the shears angles ($\theta_{1}$ and $\theta_{I}$) for the dipole band DB II.}
\end{figure}

The dipole bands in this mass region have often been interpreted as MR bands on the basis of theoretical calculations 
within the framework of the Shears mechanism with Principal Axis Cranking (SPAC) 
model \cite{podsvi, pasern1, rajban, rajban1, rajban3}. The same semiclassical calculation has been carried out
for the dipole band DB I in $^{143}$Sm, having assigned it a $\pi h_{11/2}^{4} {\otimes}$ $\nu{h}_{11/2}^{-1}$ $\pi(g_{7/2}/d_{5/2})^{-2}$ configuration. 
The Total Routhian surface (TRS) calculation \cite{naza1, naza2} exhibit a prolate minimum [$\beta_{2}$ $\approx$ 0.10 at rotational frequency ($\omega$) $\approx$ 0.200 MeV] 
associated with this configuration in $^{143}$Sm nucleus. The excitation of four proton particles into the $h_{11/2}$ orbital is favored by the  
sharp decreasing trend of its energy at the aforementioned deformation and similar excitations have previously been reported 
for $^{149}$Gd ($Z = 64, N = 85$) nucleus \cite{Fli91} with deformation of the same order.
The preferential alignment of the angular momentum vectors produced by the particle and the hole, 
under prolate deformation, is along the rotation axis and the deformation axis, respectively. 
The SPAC model, that has been elaborated elsewhere (Ref. \cite{rajban,rajban1}, for example), pertains to the minimization of the total energy of an excited state with a given spin
with respect to the angles of the two (particle and hole) angular momentum blades with rotation axis. The angles corresponding to the minimized energy are applied to determine 
the B(M1) and B(E2) values of the respective states. In the present case,  
the SPAC calculations have been performed with an unstretched condition of the hole and the particle angular momenta, $j_{1}$ = 9.5$\hbar$ and $j_{2}$ = 14$\hbar$, respectively. 
The resulting $B(M1)$, $B(E2)$ and rotational frequency ($\omega$) for the band DB I well reproduce the experimental numbers, as illustrated in Fig. 3.
The parameter values for this calculation have been adopted with reference to the similar efforts in the neighboring nuclei \cite{pasern, podsvi, pasern1, rajban1, rajban, rajban3}. 
The successful representation of the decreasing trend of experimental $B(M1)$ and $B(E2)$ values with spin as well as the behaviour of $\omega$ therein, 
within the purview of the SPAC model, is strongly indicative of the fact that the maximum contribution to the angular momentum of the states constituting DB I 
is from the shears mechanism, thus confirming the earlier interpretation by Raut {\it{et al.}} \cite{rraut}. 
However, it is intriguing to address the rapidly falling trend of the $B(M1)$/$B(E2)$ ratio  at the highest spins in DB I, so predicted by the SPAC model calculation, 
and that is in sharp contradiction to the observed sharp rise of the same (Fig. \ref{tran-bm1-be2}). 
This decreasing trend of the $B(M1)$/$B(E2)$ value might indicate a bearing from the evolving contribution of the collective core rotation along the band. 
So because there is precedence of reproducing the experimentally observed falling trend of $B(M1)$/$B(E2)$ in case of shears band in $^{139}$Sm \cite{pasern}
by incorporating an increase in the collective contribution along the sequence. Therefore, the (reverse trend of) increase in the ratio, identified in the present investigation,  
might actually be indicative of a reduction in the core contribution along the DBI band in $^{143}$Sm. \\

A confirmation to the proposition (on the evolving core contribution) was obtained from a further (semiclassical) calculation, following the prescription of the Clark and Macchiavelli \cite{rmcla1}, that was carried out for the DBI band in $^{143}$Sm, 
ascribed to the $\pi h_{11/2}^{4} {\otimes}$ $\nu{h}_{11/2}^{-1}$ $\pi(g_{7/2}/d_{5/2})^{-2}$ configuration. 
This calculation, at variance with the SPAC model, incorporates a smooth decrease in the core angular momentum ($R$) along the band as per the expression, 
$R$ = 2$\times$(26.5 - $I$)/$\xi$, $\xi$ being the number of observed states in the band. 
The scalar addition of the ($R$) with the shears angular momentum ($|\overrightarrow{j_{sh}}|$ = $|\overrightarrow{j_{p}}$ + $\overrightarrow{j_{h}}|$; 
where $\overrightarrow{j_{p}}$ and $\overrightarrow{j_{h}}$ are the angular momentum vectors for the particles and the holes, respectively) 
produces the total angular momentum of the state ($I$). The transition probabilities are calculated from the equations elaborated in Ref.~\cite{rmcla1}.
The results in the present case, carried out using $j_{p}$ = 16$\hbar$, $j_{h}$ = 10.5$\hbar$ and $\xi$ = 6, are in excellent overlap with the experimental 
$\omega$, $B(M1)$, $B(E2)$ and $B(M1)$/$B(E2)$ values in DBI, as illustrated in Figs. \ref{theo-cal} and \ref{tran-bm1-be2}. It emerges that the angular momentum 
contribution from the collective rotation, $R$, is $\sim$ 2 at the bandhead 43/2$^{-}$ and smoothly decreases to 0 at the highest observed state 53/2$^{(-)}$. Similar decreasing feature of the transition quadrupole moment parameter $Q_{t}$ has been obtained from the TRS calculation in agreement with the values derived from the present experimental results, as shown in the inset (a) of Fig. \ref{tran-bm1-be2}. This implies that the the core becomes more and more symmetric, with respect to the total angular momentum, as the angular momentum blades close-in 
to generate the high spin states in DBI and thus leads to a reduction in the collectivity along the same. Together, the SPAC and the Clark-Macchiavelli calculations
affirm the MR character of the DBI band in $^{143}$Sm along with a novel evidence of decreasing core contribution with increasing spin along the sequence. \\     

The unique observation of the present study pertains to the dipole band DB II that bifurcates from DB I at 49/2$^{-}$ and 
emerges as an energetically favoured sequence thereafter. The $I$ vs. $\omega$ plot (Fig. \ref{theo-cal}) for this band 
illustrates generation of much angular momentum ($I$) with very little increment in the rotational frequency ($\omega$), thus 
indicating an increase in the collectivity along DB II. Indeed, the SPAC calculations carried out for DB II, assigned with the same 
quasiparticle configuration as DB I, yield a satisfactory overlap with the experimental $\omega$, $B(M1)$, $B(E2)$ and $B(M1)$/$B(E2)$ values (Figs. \ref{theo-cal} and \ref{tran-bm1-be2})
only when the moment of inertia ($J$) of the core is allowed to increase with spin, as depicted in Fig. \ref{tran-bm1-be2}. This calculation 
conclusively establish that the increasing angular momentum along DB II entirely results from the increasing contribution of the core rotation therein as 
is indicated by the unit slope of $R$ vs $I$ plot in Fig. \ref{theo-cal} while the orientation of the angular momentum blades (shears) is almost fixed throughout the
sequence (Fig. \ref{tran-bm1-be2}). It may thus be stated that the DB II sequence represents a scenario where the collective core rotation
is a more favored mechanism for generation of angular momentum rather than the alignment of the angular momentum blades underlying the MR / shears phenomenon. 
It is imperative to distiguish this situation from those previously reported in $^{139, 141}$Sm \cite{pasern,rajban3} and $^{128}$Ba \cite{ovogel} wherein 
an increase in collectivity have been observed along the MR bands in sharp contrast to the present case with one of the bands (DB I) preserving its MR origin to higher spins 
while evolving into as well as coexisting with an alternative and energetically favourable band (DB II) of collective rotational character. \\      

The 49/2$^{-}$ state in the $^{143}$Sm nucleus, thus, represents a very unique transition point beyond which two radically different mechanisms
for generation of angular momentum in nuclei appear to coexist. While the states of DB I beyond 49/2$^{-}$ still originate from the aligning 
particle and hole angular momentum blades, as is the case in MR bands, the levels of increasing spin in DB II emerge from the collective rotation of the 
core. The latter being energetically favorable, the level at 49/2$^{-}$ can be perceived as a transition point from a few quasiparticle based angular momentum generation, 
as in the shears mechanism, to a domain where the collective rotation of the core is the preferred mode of excitation. Equivalently, this also implies a transformation 
from a near spherical or weakly deformed system to a well deformed one. To the best of our knowledge, this is the maiden instance when such a 
transformation has been established in the study of high-spin excited states of atomic nuclei.  \\    

In summary, high spin dipole sequences in the $^{143}$Sm nucleus have been investigated using heavy-ion induced fusion-evaporation reaction 
and an array of Compton suppressed Clover detectors. Two bands, DB I and DB II, have been confirmed from the spectroscopic measurements 
including determination of level lifetimes using DSAM. Theoretical calculations have been carried out within the purview of the SPAC model as well as
the semiclassical model of Clark and Macchiavelli. In the light of the experimental observations and their successful representation in the model calculations, 
DB I has been confirmed as a MR band while DB II, stemming out of DB I at one of its intermediate state, has been ascribed to the collective rotation of the 
core and energetically more favored compared to the corresponding same spin states of DB I. The intermediate state of DB I, with spin 49/2$^-$, has thus been identified as
a transition point from a weakly deformed system, with shears mechanism in operation for generation of angular momentum, to a deformed one, with collective rotation as the preferred 
mode for producing higher spins albeit coexisting with the MR phenomenon in the same excitation regime. This is a unique observation of such transition followed by 
coexistence in the structure studies of atomic nuclei.

\begin{center}
$\textbf{Acknowledgements}$
\end{center}

We would like to acknowledge the help from all INGA collaborators. We are thankful to the Pelletron staff for giving us steady and uninterrupted $^{24}$Mg beam. S. R. would like to acknowledge the financial assistance from the University Grants Commission - Minor Research Project (No. PSW-249/15-16 (ERO)). G. G acknowledges the support provided by the University Grants Commission - departmental research support (UGC-DRS) program. H. P. is grateful for the support of the Ramanujan Fellowsship research grant under SERB-DST (SB/S2/RJN-031/2016).



\begin{thebibliography}{}

\bibitem{abohr} A. Bohr and B. R. Mottelson, Phys. Rev. {\bf 90} (1953) 717.

\bibitem{defor1} S. Raman, C. W. Nestor, Jr, P. Tikkanen, At. Data Nucl. Data 
 Tables {\bf 78} (2001) 1.

\bibitem{defor2} N. J. Stone, At. Data Nucl. Data Tables {\bf 90} (2005) 75.

\bibitem{frauen1} S. Frauendorf, Rev. Mod. Phy. {\bf 73} (2001) 463.


\bibitem{ajsim} A. J. Simons \textit{et al.}, Phys. Rev. Lett. {\bf 91} (2003) 162501.


\bibitem{hubel} H. H$\ddot{u}$bel, Prog. Part. Nucl. Phys. {\bf 54} (2005) 1.


\bibitem{amita} Amita, A. K. Jain, and B. Singh, At. Data Nucl. Data Tables {\bf74} (2000) 283.


\bibitem{rmcla1} R. M. Clark and A. O. Macchiavelli, Annu. Rev. Nucl. Part. Sci. {\bf 50} (2000) 1.

\bibitem{pasern} A. A. Pasternak, \textit{et al.}, Eur. Phys. J. {\bf A 37} (2008) 279 - 286.

\bibitem{podsvi} E. O. Podsvirova \textit{et al.}, Eur. Phys. J.
 {\bf A 21} (2004) 1 - 6.

\bibitem{pasern1} A. A. Pasternak \textit{et al.}, Eur. Phys. J.
 {\bf A 23} (2005) 191 - 196.


\bibitem{mgproc} M. G. Procter \textit{et al.}, Phys. Rev. C {\bf 81} (2010) 054320. 

\bibitem{rraut} R. Raut \textit{et al.}, Phys. Rev. C {\bf 73} (2006) 044305.


\bibitem{smurali} S. Muralithar \textit{et al.}, Nucl. Instrum. Methods Phys. Res. {\bf A} 622 (2010) 281.

\bibitem{ingasort} R. K. Bhowmik, Ingasort Manual, private communication. 

\bibitem{radford1} D. C. Radford, Nucl. Instrum. Methods, Phys. Res.,
 Sect {\bf A 361} (1995) 297.

\bibitem{radford2} D. C. Radford, Nucl. Instrum. Methods, Phys. Res.,
 Sect {\bf A 361} (1995) 306. 




\bibitem{kramer} A. Kr$\ddot{\textup{a}}$mer-Flecken \textit{et al.}, Nucl. Instrum. Methods. phys. res. {\bf A 275} (1989) 333-339.

\bibitem{kabadi} M.K. Kabadiyski, K.P. Lieb, D. Rudolph, Nucl. Phys. {\bf A 563} (1993) 301-325.


\bibitem{staro2} K. Starosta \textit{et al.}, Nucl. Instrum. Methods, Phys. Res. A {\bf 423} (1999) 16 - 26.

\bibitem{droste2} Ch. Droste \textit{et al.}, Nucl. Instrum. Methods Phys. Res. A {\bf 378} (1996) 518-525.

\bibitem{deng2} J. K. Deng \textit{et al.}, Nucl. Instrum. Methods Phys. Res. A {\bf 317} (1992) 242.

\bibitem{jones2} P. M. Jones \textit{et al.}, Nucl. Instrum. Methods Phys. Res. A {\bf 362} (1995) 556.

\bibitem{tyamaza} T. Yamazaki, Nucl. Data section A {\bf 3} (1967) 1.

\bibitem{edmawsu} E.D. Mateosion and A.W. Sunyar, Atomic Data. Nucl. Data Tables {\bf 13} (1974) 392.
 

\bibitem{macias} E. S. Macias, W. D. Ruhter, D.C. Camp and R.G. Lanier, Computer Physics Communications {\bf 11} (1976) 75—93.


\bibitem{rajban1} S. Rajbanshi \textit{et al.}, Phys. Rev. C {\bf 90} (2014) 024318.


\bibitem{rajban} S. Rajbanshi \textit{et al.}, Phys. Rev. C {\bf 89} (2014) 014315.

\bibitem{das16} S. Das \textit{et al.}, Nucl. Instrum. Methods Phys. Res. A {\bf 841} (2017) 17.

\bibitem{wells} J.C. Wells, N.R. Johnson, LINESHAPE: A Computer Program 
for Doppler Broadened Lineshape Analysis, Report No. ORNL-6689  (1991) 44.

\bibitem{johnson} N.R. Johnson \textit{et al.}, Phys. Rev. C {\bf 55} (1997) 652.




\bibitem{rschwen} R. Schwengner \textit{et al.}, Phys. Rev. C {\bf 80} (2009) 044305.


\bibitem{rajban3} S. Rajbanshi \textit{et al.}, Phys. Rev. C {\bf 94} (2016) 044318.



\bibitem{naza1} W. Nazarewicz, J. Dudek, R. Bengtsson, T. Bengtsson, and I. Ragnarsson, Nucl. Phys. {\bf A 435} (1985) 397.

\bibitem{naza2} W. Nazarewicz, M. A. Riley, and J. D. Garrett, Nucl. Phys. {\bf A 512} (1990) 61.


\bibitem{ovogel} O. Vogel  \textit{et al.}, Phys. Rev. C {\bf 56} (1997) 1338.



\bibitem{Fli91} S. Flibotte  \textit{et al.}, Nucl. Phy. A {\bf 530} (1991) 187.






\end{thebibliography}
\end{document}